\documentclass[12pt]{article}
\usepackage{amsmath,amsfonts,amssymb}

\begin{document}
\thispagestyle{empty}

\def\theequation{\arabic{section}.\arabic{equation}}
\def\a{\alpha}
\def\b{\beta}
\def\g{\gamma}
\def\d{\delta}
\def\dd{\rm d}
\def\e{\epsilon}
\def\ve{\varepsilon}
\def\z{\zeta}
\def\B{\mbox{\bf B}}\def\cp{\mathbb {CP}^3}

\newcommand{\h}{\hspace{0.5cm}}

\begin{titlepage}
\vspace*{1.cm}
\renewcommand{\thefootnote}{\fnsymbol{footnote}}
\begin{center}
{\Large \bf Three-point Correlation functions of}
\vskip 0cm
{\Large\bf Giant magnons with finite size}
\end{center}
\vskip 1.2cm \centerline{\bf Changrim  Ahn and Plamen Bozhilov
\footnote{On leave from Institute for Nuclear Research and Nuclear
Energy, Bulgarian Academy of Sciences, Bulgaria.}}

\vskip 10mm

\centerline{\sl Institute for the Early Universe and Department of Physics} \centerline{\sl Ewha Womans
University} \centerline{\sl DaeHyun 11-1, Seoul 120-750, S. Korea}
\vspace*{0.6cm} \centerline{\tt ahn@ewha.ac.kr,
bozhilov@inrne.bas.bg}

\vskip 20mm

\baselineskip 18pt

\begin{center}
{\bf Abstract}
\end{center}

We compute holographic three-point correlation functions or structure constants
of a zero-momentum dilaton operator and two (dyonic)
giant magnon string states with a finite-size length in the semiclassical approximation.
We show that the semiclassical structure constants match exactly with the three-point functions between two
$su(2)$ magnon single trace operators with finite size and the Lagrangian in the large 't Hooft coupling constant limit.
A special limit $J\gg\sqrt{\lambda}$ of our result is compared with the relevant
result based on the L\"uscher corrections.

\end{titlepage}
\newpage
\baselineskip 18pt

\def\nn{\nonumber}
\def\tr{{\rm tr}\,}
\def\p{\partial}
\newcommand{\non}{\nonumber}
\newcommand{\bea}{\begin{eqnarray}}
\newcommand{\eea}{\end{eqnarray}}
\newcommand{\bde}{{\bf e}}
\renewcommand{\thefootnote}{\fnsymbol{footnote}}
\newcommand{\be}{\begin{eqnarray}}
\newcommand{\ee}{\end{eqnarray}}

\vskip 0cm

\renewcommand{\thefootnote}{\arabic{footnote}}
\setcounter{footnote}{0}

\setcounter{equation}{0}
\section{Introduction}
Correlation functions of conformal field theories (CFTs) can be in principle
determined in terms of basic conformal data $\{\Delta_i,C_{ijk}\}$,
where $\Delta_i$'s are
conformal dimensions defined by two-point correlation functions
\begin{equation}
\left\langle{\cal O}^{\dagger}_i(x_1){\cal O}_j(x_2)\right\rangle=
\frac{\delta_{ij}}{|x_1-x_2|^{2\Delta_i}}
\end{equation}
and $C_{ijk}$'s are structure constants by three-point correlation functions
\begin{equation}
\left\langle{\cal O}_i(x_1){\cal O}_j(x_2){\cal O}_k(x_3)\right\rangle=
\frac{C_{ijk}}{|x_1-x_2|^{\Delta_1+\Delta_2-\Delta_3}
|x_1-x_3|^{\Delta_1+\Delta_3-\Delta_2}|x_2-x_3|^{\Delta_2+\Delta_3-\Delta_1}}.
\end{equation}
Hence a complete determination of conformal data for a given CFT is a most
important step in the conformal bootstrap procedure.
While this is well-established in two dimensions where the conformal symmetry is
infinite dimensional \cite{BPZ},
it is extremely difficult to extend the precedure to higher space-time dimensions.

In four dimensions, the AdS/CFT correspondence between the ${\cal N}=4$ super
Yang-Mills theory (SYM) and type IIB string theory moving on ${\rm AdS}_5\times S^5$
target space has provided a most promising framework \cite{AdS/CFT}.
A lot of impressive progresses have been made in this field based on the integrability
discovered in the planar limit of the SYM.
In particular, the thermodynamic Bethe ansatz approach based on non-perturbative
world-sheet $S$-matrix has been formulated to provide the conformal dimensions of
SYM operators with arbitrary number of elementary fields for
generic value of 't Hooft coupling constant $\lambda$
(see for a recent review \cite{review}).
In strong coupling limit $\lambda\gg 1$ the AdS/CFT correspondence relates the
conformal dimensions to energy of certain classical string configurations which
can be computed by either solving the superstring sigma model directly such as
the algebraic curve method \cite{SpecCurv} or Neumann-Rosochatius reduction \cite{KRT}.

There have been many interesting progresses on three-point correlation functions
in the AdS/CFT context.
Three-point functions for chiral primary operators have been
computed first in the ${\rm AdS}_5$ supergravity approximation where explicit dependence on the coupling constant $\lambda$ is not apparent \cite{sugra}.
It is only recently that several interesting developments have been made to consider general heavy string states.
An efficient method to compute two-point correlation functions in the strong coupling limit is to evaluate string
partition function for a heavy string state propagating in the AdS space between two boundary points based on a path integral method \cite{Janik,BuchTsey}.
This method has been extended to the three-point functions of two heavy string states
and a supergravity mode which corresponds to a marginal deformation of the SYM
two-point functions by the Lagrangian \cite{Zarembo,Costa,Roiban}.
With these formulation, many interesting checks of three-point functions of two heavy mode and one light mode
have been performed \cite{Hernandez}-\cite{Hernandez2}.
Another direction is to compute the structure constants using the Bethe eigenstates of the underlying integrable
spin chain in the weak coupling limit of the SYM \cite{Gromov,Vieira}.

In this note we apply the semiclassical formulation of the three-point correlation functions of a ``zero-momentum'' dilaton operator which is the Lagrangian as a light operator along with two heavy (dyonic) giant magnon string states.
Differently from previous cases with an infinite length of the SYM operator $J\to\infty$ \cite{Costa,Hernandez2},
we consider finite-size systems $J\sim \sqrt{\lambda}\gg 1$.
We show that the semiclassical formulation of the three-point functions is still valid for this more general situation.
A special limit of our result is $J\gg \sqrt{\lambda}$ where the finite-size corrections to both conformal dimensions and
energies of string states have been
computed from the L\"uscher corrections.

The three-point functions of two heavy operators and a light operator
can be approximated by a supergravity vertex operator evaluated at the heavy
classical string configuration;
\bea
\nn
\langle V_{H_1}(x_1)V_{H_2}(x_2)V_{L}(x_3)\rangle=V_L(x_3)_{\rm classical}.
\eea
For $\vert x_1\vert=\vert x_1\vert=1$, $x_3=0$, the correlation function reduces to
\bea
\nn
\langle V_{H_1}(x_1)V_{H_2}(x_2)V_{L}(0)\rangle=\frac{C_{123}}{\vert
x_1-x_2\vert^{2\Delta_{H_1}}}.
\eea
Then, the normalized structure constants
\bea
\nn
\mathcal{C}_3=\frac{C_{123}}{C_{12}}
\eea
can be found from \cite{Hernandez2}
\bea
\label{nsc}
\mathcal{C}_3=c_{\Delta}V_L(0)_{\rm classical}.
\eea
Here, $c_{\Delta}$ is the normalized constant of the corresponding light vertex
operator.

We restrict our consideration to the zero-momentum dilaton operator, namely the Lagrangian whose vertex operator is given by 
\bea \label{dv}
V^d=\left(Y_4+Y_5\right)^{-4}\left[z^{-2}\left(\p_+x_{m}\p_-x^{m}+\p_+z\p_-z\right)
+\p_+X_{k}\p_-X_{k}\right], \eea where \bea \nn
Y_4=\frac{1}{2z}\left(x^mx_m+z^2-1\right), \h
Y_5=\frac{1}{2z}\left(x^mx_m+z^2+1\right), \eea and $x_m$, $z$ are
coordinates on $AdS_5$, while $X_k$ are the coordinates on $S^5$.

\setcounter{equation}{0}
\section{Giant magnons with finite size}
The finite-size giant magnon solution \cite{AFZ06,AFGS,KM0803}, in
the notations of \cite{AB1} can be
represented as ($i\tau=\tau_e$) \bea\nn
&&x_{0e}=\tanh(\kappa\tau_e),\h x_i=0,\h
z=\frac{1}{\cosh(\kappa\tau_e)},
\\ \label{fsgm} &&\cos\theta=\sqrt{1-v^2\kappa^2}
dn\left(\frac{\sqrt{1-v^2\kappa^2}}{1-v^2}(\sigma-v\tau)\Big\vert
1-\epsilon\right) ,
\\ \nn &&\phi= \frac{\tau-v\sigma}{1-v^2}+\frac{1}{v\sqrt{1-v^2\kappa^2}}\times
\\ \nn &&
\Pi\left(-\frac{1-v^2\kappa^2}{v^2\kappa^2}\left(1-\epsilon\right),
am\left(\frac{\sqrt{1-v^2\kappa^2}}{1-v^2}(\sigma-v\tau)\right)\Big\vert
1-\epsilon\right),\eea where $dn(x\vert y)$
is one of the Jacobi elliptic functions, $am(x)$ is the Jacobi
amplitude, and $\Pi(x,y\vert z)$ is the incomplete elliptic integral
of third kind, and 
\bea\nn \epsilon=\frac{1-\kappa^2}{1-v^2\kappa^2}.\eea

To find the finite-size effect on the three-point correlator, we
will use (\ref{nsc}) and (\ref{dv}), which computed on (\ref{fsgm})
gives \bea\label{c3d}
\mathcal{C}_3^d=c_{\Delta}^{d}\int_{-\infty}^{\infty}\frac{d\tau_e}{\cosh^4(\kappa\tau_e)}
\int_{-L}^{L}d\sigma\left[\kappa^2 +\p_+X_{k}\p_-X_{k}\right],\eea
where\bea\nn \p_+X_{k}\p_-X_{k}&=&-\frac{1}{(1-v^2)\sin^2\theta}
\left\{2-(1+v^2)\kappa^2\right.
\\ \nn &&-\left.\cos^2\theta\left[4-(1+v^2)\kappa^2
-2\cos^2\theta\right]\right\} .\eea
The integration variable $\sigma$ can be changed to $\xi=\sigma-v\tau$ and 
to $\theta$,
\bea
\int^L_{-L}d\sigma\ldots =
\int^{L-v\tau}_{-L-v\tau}d\xi\ldots
=2\int^{\theta_{\rm max}}_{\theta_{\rm min}}\frac{d\theta}{\theta'(\xi)}\ldots
\eea
using the periodic property of $\theta(\xi)$ 
(periodicity is $2\mathbf{K}(1-\epsilon)$) and the integral
becomes independent of $\tau$.
We would like to emphasize that, as explained in \cite{AFZ06},
unlike the infinite size giant magnon \cite{HM06}, the finite-size
giant magnon is {\it non-rigid}. 
When $L\to\infty$, the string becomes rigid and the end points touch the equator.

Performing the integrations in (\ref{c3d}), one finds
\bea\label{exact1} \mathcal{C}_3^d=\frac{16}{3}c_{\Delta}^{d}
\sqrt{\frac{1-v^2}{1-\epsilon}}
\left[\mathbf{E}(1-\epsilon)-\epsilon\mathbf{K}(1-\epsilon)\right],\eea
where $\mathbf{K}(1-\epsilon)$ and $\mathbf{E}(1-\epsilon)$ are the
complete elliptic integrals of first and second kind. 
Let us point out that the parameter $L$ in (\ref{exact1}) is given
by \bea\nn
L=\frac{1-v^2}{\sqrt{1-v^2\kappa^2}}\mathbf{K}(1-\epsilon).\eea This
is our {\it exact} result for the normalized coefficient
$\mathcal{C}_3^d$ in the three-point correlation function,
corresponding to the case when the heavy vertex operators are {\it
finite-size} giant magnons, and the light vertex is taken to be the
zero-momentum dilaton operator.

For the case of this dilaton operator, 
the three-point function of the SYM can be
easily related to the conformal dimension of the heavy operators.
This corresponds to shift `t Hooft coupling constant which
is the overall coefficient of the Lagrangian \cite{Costa}.
This gives an important relation between the structure constant and the conformal
dimension as follows:
\bea\label{rel1}
\mathcal{C}_3^d=\frac{32\pi}{3}c_{\Delta}^{d}\sqrt{\lambda}\p_\lambda\Delta.
\eea
We want to show that this relation is correct for the case of the giant magnons
with arbitrary finite size.

In the context of the AdS/CFT correspondence, it is now well-established that
the conformal dimension of a single trace operator with one magnon state is the
same as $E-J$ in the strong coupling limit.
For an exact relation from the gauge theory side, one should solve the
thermodynamic Bethe equations \cite{TBA}.
Although this is very complicated and analytic solutions are still not available,
it has been shown that finite-size corrections to the conformal dimensions of
the SYM (dyonic) giant magnon operators computed by 
the L\"uscher formula for $J\gg\sqrt{\lambda}$ 
match exactly with $E-J$ of corresponding
string state configurations \cite{JanLuk,HatSuz}.
Based on these results, we can assume that the conformal dimensions
$\Delta$ of the
SYM operators are the same as $E-J$ of corresponding string states.

The {\it exact} classical expression for finite-size giant magnon
energy-charge relation is given by \cite{AB1} \bea\label{eEJ}
E-J\equiv \Delta=
\frac{\sqrt{\lambda}}{\pi} \sqrt{\frac{1-v^2}{1-v^2\epsilon}}
\left[\mathbf{E}(1-\epsilon)-\left(1-\sqrt{(1-v^2\epsilon)
(1-\epsilon)}\right)\mathbf{K}(1-\epsilon)\right]. \eea The
corresponding expressions for $J$ and $p$ are \bea\nn
&&J=\frac{\sqrt{\lambda}}{\pi}\sqrt{\frac{1-v^2}{1-v^2\epsilon}}
\left[\mathbf{K}(1-\epsilon)-\mathbf{E}(1-\epsilon)\right],\\ \nn
&&p=2v\sqrt{\frac{1-v^2\epsilon}{1-v^2}}
\left[\frac{1}{v^2}\Pi\left(1-\frac{1}{v^2}\Big\vert
1-\epsilon\right)-\mathbf{K}(1-\epsilon)\right],\eea where
$J$ is the angular momentum
of the string, and $p$ is the magnon momentum. One can obtain $E-J$
in terms of $J$ and $p$ by eliminating $v, \epsilon$ from these
expressions.

To take $\lambda$-derivative on $\Delta$,
we need know $\lambda$ dependence of $v$ and $\epsilon$.
Our strategy is to find $v'(\lambda)$ and $\epsilon'(\lambda)$ from
the conditions that $J$ and $p$ are independent variables of $\lambda$,
namely,
\bea\label{eqs} \frac{dJ}{d\lambda}=\frac{dp}{d\lambda}=0.\eea 
Solving these conditions, we find the derivatives of
the functions $v(\lambda)$ and $\epsilon(\lambda)$ 
\bea\label{spd}
&&\frac{dv}{d\lambda}= -\frac{v
(1-v^2)\epsilon\left[\mathbf{E}(1-\epsilon)-\mathbf{K}(1-\epsilon)\right]^{2}}
{2\lambda(1-\epsilon)\left[\mathbf{E}(1-\epsilon)^2
-v^2\epsilon\mathbf{K}(1-\epsilon)^2\right]}, \\ \nn
&&\frac{d\epsilon}{d\lambda}=-
\frac{\epsilon\left[\mathbf{E}(1-\epsilon)-\mathbf{K}(1-\epsilon)\right]
\left[\mathbf{E}(1-\epsilon)-v^2\epsilon\mathbf{K}(1-\epsilon)\right]}
{\lambda\left[\mathbf{E}(1-\epsilon)^2
-v^2\epsilon\mathbf{K}(1-\epsilon)^2\right]}.\eea Replacing
(\ref{spd}) into the derivative of (\ref{eEJ}), one finds
\bea\label{ddr}
\lambda\p_\lambda\Delta=\frac{\sqrt{\lambda}}{2\pi}
\sqrt{\frac{1-v^2}{1-\epsilon}}
\left[\mathbf{E}(1-\epsilon)-\epsilon\mathbf{K}(1-\epsilon)\right].\eea
Comparing (\ref{exact1}) and (\ref{ddr}), we conclude that the
equality (\ref{rel1}) holds.

Next, we would like to compare (\ref{exact1}) with the known leading
finite-size correction to the giant magnon dispersion relation
\cite{AFZ06}. To this end, we have to consider the limit
$\epsilon\to 0$ in (\ref{exact1}). Taking into account the behavior
of the elliptic integrals in the $\epsilon\to 0$ limit, we can use
the ansatz \bea\label{vexp} v(\epsilon)=v_0+v_1\epsilon +
v_2\epsilon\log(\epsilon).\eea Actually, all parameters in
(\ref{vexp}) are already known and are given by (see for instance
\cite{AB1}) \bea\label{vsol} &&v_0=\cos(p/2),\h
v_1=\frac{1}{4}\sin^2(p/2)\cos(p/2)(1-\log(16)),\\ \nn &&
v_2=\frac{1}{4}\sin^2(p/2)\cos(p/2),\h\epsilon=16
\exp{\left(-\frac{2\pi J}{\sqrt{\lambda}\sin(p/2)}-2\right)}.\eea  
Expanding
(\ref{exact1}) in $\epsilon$ and using (\ref{vexp}), (\ref{vsol}),
we obtain \bea\label{Cexp}
\mathcal{C}_3^d=\frac{16}{3}c_{\Delta}^{d}\sin(p/2)
\left[1-4\sin(p/2)\left(\sin(p/2)+
\frac{2\pi J}{\sqrt{\lambda}}\right)\exp{\left(-\frac{2\pi J}{\sqrt{\lambda}\sin(p/2)}-2\right)}\right].\eea

On the other hand, from the giant magnon dispersion relation,
including the leading finite-size effect,  \bea\nn
\Delta=\frac{\sqrt{\lambda}}{\pi}\sin(p/2)\left[1-4\sin^2(p/2)
\exp{\left(-\frac{2\pi
J}{\sqrt{\lambda}\sin(p/2)}-2\right)}\right],\eea one finds
\bea\label{der1} \lambda\p_\lambda\Delta=\frac{\sqrt{\lambda}}{2\pi}
\sin(p/2) \left[1-4\sin(p/2)\left(\sin(p/2)+
\frac{2\pi J}{\sqrt{\lambda}}\right)\exp{\left(-\frac{2\pi J}{\sqrt{\lambda}\sin(p/2)}-2\right)}\right].\eea
This confirms explicitly that the relation (\ref{rel1}) holds in the
small $\epsilon$ i.e. $J\gg\sqrt{\lambda}$ limit.

\setcounter{equation}{0}
\section{Dyonic giant magnons with finite size}

The dyonic finite-size giant magnon solution is given by \bea\nn
&&x_{0e}=\tanh(\kappa\tau_e),\h x_i=0,\h
z=\frac{1}{\cosh(\kappa\tau_e)},
\\ \label{fsgm2} &&\cos\theta=z_+
dn\left(\frac{\sqrt{1-u^2}}{1-v^2}z_+(\sigma-v\tau)\Big\vert
1-\epsilon\right) ,
\\ \nn &&\phi_1= \frac{\tau-v\sigma}{1-v^2}+\frac{vW}{\sqrt{1-u^2}z_+(1-z_+^2)}\times
\\ \nn &&
\Pi\left(-\frac{z_+^2}{1-z_+^2}\left(1-\epsilon\right),
am\left(\frac{\sqrt{1-u^2}}{1-v^2}z_+(\sigma-v\tau)\right)\Big\vert
1-\epsilon\right) \\ \nn &&\phi_2=
u\frac{\tau-v\sigma}{1-v^2},\eea where \bea\nn
\epsilon=\frac{z_-^2}{z_+^2},\h W=\kappa^2.\eea $z_\pm^2$ can be
written as \bea\nn &&z^2_\pm=\frac{1}{2(1-u^2)} \left\{q_1+q_2-u^2
\pm\sqrt{(q_1-q_2)^2-\left[2\left(q_1+q_2-2q_1 q_2\right)-u^2\right]
u^2}\right\}, \\ \nn &&q_1=1-W,\h q_2=1-v^2W .\eea

Now, we have to replace into (\ref{c3d}) the following expression
obtained from the above solution \bea\nn
\p_+X_{k}\p_-X_{k}&=&-\frac{1}{(1-v^2)\sin^2\theta}
\left\{1-v^2W^2+(1-u^2)z_+^4\epsilon +2(1-u^2)\cos^4\theta\right. \\
\nn &&-
\left.\cos^2\theta\left[2+z_+^2(1+\epsilon)-u^2\left(1+z_+^2(1+\epsilon)\right)\right]\right\}
.\eea Computing the integrals in (\ref{c3d}), we find
\bea\label{exact2} &&\mathcal{C}_3^d=\frac{8}{3}c_{\Delta}^{d}
\frac{1}{\sqrt{(1-u^2)W\chi_p}(1-\chi_p)}
\Bigg\{(1-\chi_p)\left[2(1-u^2)\chi_p\mathbf{E}(1-\epsilon)\right.
\\ \nn &&
-\left.\left(u^2-\left(1-v^2\right)W+(1-u^2)(1+\epsilon)\chi_p\right)\mathbf{K}(1-\epsilon)\right]
\\ \nn &&
-\left(1-v^2W^2-\chi_p-(1-\chi_p)\left(\epsilon\chi_p
+u^2(1-\epsilon\chi_p)\right)\right)\times
\\ \nn &&
\Pi\left(-\frac{\chi_p}{1-\chi_p}\left(1-\epsilon\right)\Big\vert
1-\epsilon\right)\Bigg\},\eea where we introduced the notations
\bea\nn \chi_p=z_+^2,\h \chi_m=z_-^2,\h \Rightarrow
\epsilon=\frac{\chi_m}{\chi_p}.\eea This is our {\it exact} result
for the normalized coefficient $\mathcal{C}_3^d$ in the three-point
correlation function, corresponding to the case when the heavy
vertex operators are {\it finite-size} dyonic giant magnons.

To check the relation (\ref{rel1}), we need to know $\Delta$.
As GM case, we claim that this is given by $E-J_1$.
The explicit results are given by \cite{Mpaper}
\bea\nn &&\mathcal{E}
=\frac{2\sqrt{W}(1-v^2)} {\sqrt{1-u^2}\sqrt{\chi_p}}\mathbf{K}
\left(1-\epsilon\right), \\ \label{cqsGM} &&\mathcal{J}_1=
\frac{2\sqrt{\chi_p}} {\sqrt{1-u^2}}\left[
\frac{1-v^2W}{\chi_p}\mathbf{K}
\left(1-\epsilon\right)-\mathbf{E}
\left(1-\epsilon\right)\right], \\ \nn &&\mathcal{J}_2=
\frac{2u\sqrt{\chi_p}} {\sqrt{1-u^2}}\mathbf{E}
\left(1-\epsilon\right)\\
&&p=\frac{2v} {\sqrt{1-u^2}\sqrt{\chi_p}}
\left[\frac{W}{1-\chi_p}\Pi\left(-\frac{\chi_p}{1-\chi_p}(1-\epsilon)\bigg\vert 1-\epsilon\right) -\mathbf{K}
\left(1-\epsilon\right)\right],
\eea 
and \bea\nn  \mathcal{E}=\frac{2\pi
E}{\sqrt{\lambda}},\qquad\mathcal{J}_{1,2}=\frac{2\pi
J_{1,2}}{\sqrt{\lambda}}.\eea

In this case, we need to obtain $v'(\lambda), \epsilon'(\lambda), u'(\lambda)$ from
the condition that $J_1, J_2, p$ be independent of $\lambda$. 
It turns out that the exact calculations for these are too complicated.
Instead, we will just focus on the $\epsilon\to 0$ limit of
(\ref{exact2}) and $\lambda$ derivative of $\Delta$ from the L\"uscher formula
to check (\ref{rel1}). 
To this end, we will use the
expansions \bea\nn
&&\chi_p=\chi_{p0}+\left(\chi_{p1}+\chi_{p2}\log(\epsilon)\right)\epsilon,
\h\chi_m=\chi_{m1}\epsilon,
\\
\label{Dpars} &&v=v_0+\left(v_1+v_2\log(\epsilon)\right)\epsilon, \h
u=u_0+\left(u_1+u_2\log(\epsilon)\right)\epsilon,
\\ \nn &&W=1+W_{1}\epsilon
.\eea

First note that $\chi_p$ and $\chi_m$ satisfy the following
relations \bea\label{chirel} &&\chi_p+\chi_m= \frac{2-(1+v^2)W-u^2}{1-u^2}\\
\nn && \chi_p\chi_m=\frac{1-(1+v^2)W-v^2W^2}{1-u^2}.\eea Expanding
(\ref{chirel}) and using the definition of $\epsilon$, we arrive at
\bea\label{chi} &&\chi_{p0}=1-\frac{v_0^2}{1-u_0^2}, \\
\nn &&\chi_{p1}=
\frac{v_0}{\left(1-v_0^2\right)\left(1-u_0^2\right)^2}
\Big\{v_0\left[(1-v_0^2)^2-3(1-v_0^2)u_0^2+2u_0^4-2(1-v_0^2)u_0u_1\right]
\\ \nn
 &&\h\h -2\left(1-v_0^2\right)\left(1-u_0^2\right)v_1\Big\},\\ \nn
&&\chi_{p2}=
-2v_0\frac{v_2+(v_0u_2-u_0v_2)u_0}{\left(1-u_0^2\right)^2} \\
\nn &&\chi_{m1}=1-\frac{v_0^2}{1-u_0^2},
\\ \nn &&W_1=-\frac{(1-u_0^2-v_0^2)^2}
{(1-u_0^2)(1-v_0^2)}.\eea

The coefficients in the expansions of $v$ and $u$, we take from
\cite{PB10}, where for the case under consideration we have to set
$K_1=\chi_{n1}=0$, or equivalently $\Phi=0$. This gives
\bea\label{zms}
&&v_0=\frac{\sin(p)}{\sqrt{\mathcal{J}_2^2+4\sin^2(p/2)}},\h
u_0=\frac{\mathcal{J}_2}{\sqrt{\mathcal{J}_2^2+4\sin^2(p/2)}}
\\ \nn &&v_1=\frac{v_0(1-v_0^2-u_0^2)}{4(1-u_0^2)(1-v_0^2)} \left[(1-v_0^2)(1-\log(16))
-u_0^2\left(5-v_0^2(1+\log(16))-\log(4096)\right)\right]
\\ \nn &&v_2=\frac{v_0(1-v_0^2-u_0^2)}{4(1-u_0^2)(1-v_0^2)} \left[1-v_0^2-u_0^2(3+v_0^2)\right]
\\ \nn &&u_1=\frac{u_0(1-v_0^2-u_0^2)}{4(1-v_0^2)} \left[1-\log(16)-v_0^2(1+\log(16))\right]
\\ \nn &&u_2= \frac{u_0(1-v_0^2-u_0^2)}{4(1-v_0^2)} (1+v_0^2).\eea

We need also the expression for $\epsilon$. To the leading order, it
can be written as \cite{PB10} \bea\label{eps}
\epsilon=16\exp\left(-\frac{2-\frac{2
v_0^2}{1-u_0^2}+\mathcal{J}_1\sqrt{1-v_0^2-u_0^2}}{1-v_0^2}\right).\eea

By using (\ref{chi}), (\ref{zms}) and (\ref{eps}) in the
$\epsilon$-expansion of (\ref{exact2}), we derive \bea\label{exp2}
&&\mathcal{C}_3^d=\frac{16}{3}c_{\Delta}^{d}
\Bigg\{\frac{\mathcal{J}_2^2+4\sin^2(p/2)-16\sin^4(p/2)\exp(f)}{2\sqrt{\mathcal{J}_2^2+4\sin^2(p/2)}}
\\ \nn &&
+\frac{1}{\left(\mathcal{J}_2^2+4\sin^2(p/2)\right)\left(\mathcal{J}_2^2+4\sin^4(p/2)\right)^2}
\left[32\exp(f)\left(2\mathcal{J}_2^2\sqrt{\mathcal{J}_2^2+4\sin^2(p/2)}-3\mathcal{J}_1\right.\right.
\\ \nn &&
+2\left.\left.\left(\mathcal{J}_1\left(2+\mathcal{J}_2^2\right)
+\mathcal{J}_2^2
\sqrt{\mathcal{J}_2^2+4\sin^2(p/2)}\right)\cos(p)-\mathcal{J}_1\cos(2p)\right)\sin^8(p/2)\right]
\\ \nn &&-\frac{\mathcal{J}_2^2}{2\sqrt{\mathcal{J}_2^2+4\sin^2(p/2)}}
-\frac{8\mathcal{J}_2^2\sin^4(p/2)}{\left(\mathcal{J}_2^2+4\sin^2(p/2)\right)^{3/2}}\exp(f)\Bigg\},\eea
where \bea\nn f= -\frac{2\left(\mathcal{J}_1 +
\sqrt{\mathcal{J}_2^2+4\sin^2(p/2)}\right)
\sqrt{\mathcal{J}_2^2+4\sin^2(p/2)}\sin^2(p/2)}{\mathcal{J}_2^2+4\sin^4(p/2)}
.\eea 

On the other hand, from the dyonic giant magnon dispersion relation,
including the leading finite-size correction, \bea\label{dfsdr}
\Delta_{dyonic}=\frac{\sqrt{\lambda}}{2\pi}\left[
\sqrt{\mathcal{J}_2^2+4\sin^2(p/2)} - \frac{16 \sin^4(p/2)}
{\sqrt{\mathcal{J}_2^2+4\sin^2(p/2)}}\exp(f)\right],\eea one obtains
\bea\label{der2} &&\lambda\p_\lambda\Delta_{dyonic}=
\frac{\sqrt{\lambda}}{2\pi}\Bigg\{\frac{\mathcal{J}_2^2+4\sin^2(p/2)-16\sin^4(p/2)\exp(f)}{2\sqrt{\mathcal{J}_2^2+4\sin^2(p/2)}}
\\ \nn &&
+\frac{1}{\left(\mathcal{J}_2^2+4\sin^2(p/2)\right)\left(\mathcal{J}_2^2+4\sin^4(p/2)\right)^2}
\left[32\exp(f)\left(2\mathcal{J}_2^2\sqrt{\mathcal{J}_2^2+4\sin^2(p/2)}-3\mathcal{J}_1\right.\right.
\\ \nn &&
+2\left.\left.\left(\mathcal{J}_1\left(2+\mathcal{J}_2^2\right)
+\mathcal{J}_2^2
\sqrt{\mathcal{J}_2^2+4\sin^2(p/2)}\right)\cos(p)-\mathcal{J}_1\cos(2p)\right)\sin^8(p/2)\right]
\\ \nn &&-\frac{\mathcal{J}_2^2}{2\sqrt{\mathcal{J}_2^2+4\sin^2(p/2)}}
-\frac{8\mathcal{J}_2^2\sin^4(p/2)}{\left(\mathcal{J}_2^2+4\sin^2(p/2)\right)^{3/2}}\exp(f)\Bigg\}.\eea
Comparing (\ref{exp2}) and (\ref{der2}), we see that the relation
(\ref{rel1}) is also valid for finite-size dyonic giant magnons, as
it should be.

\setcounter{equation}{0}
\section{Concluding Remarks}
In this note we have considered three-point correlation functions in the
strong coupling side of the AdS/CFT correpondence.
We have used a formulation for semiclassical structure constants of
a zero-momentum dilaton operator and two heavy string states  of (dyonic)
giant magnons of finite-size and showed that they match with the expected results
coming from derivatives of two-point correlation functions w.r.t. 't Hooft
coupling constant

It is still not clear how to overcome the key approximations we and many other
related papers have assumed.
It should be essential to utilize the integrability to consider correlation
functions for arbitrary value of 't Hooft coupling constant.
Developments in this line have been recently reported in \cite{Gromov,Vieira}.
Another issue is to include other light operators such as generic dilaton 
operators which is dual to the SYM or even general heavy string states.

Considering the remarkable developments on the two-point functions, the semiclassical anlaysis has made crucial contributions in figuring out exact integrability structure
hidden in the AdS/CFT correspondence.
We hope that our semiclassical results for the generic finite-size operators can be
a small starting point toward exact formulation of the three-point correlation functions.

\section*{Acknowledgements}
We thank Chanyong Park for useful discussions.
This work was supported in part by WCU Grant No. R32-2008-000-101300
(C. A.) and DO 02-257 (P. B.).


\begin{thebibliography}{99}
\bibitem{BPZ} A. A. Belavin, A. M. Polyakov and A. B . Zamolodchikov,
``Infinite conformal symmetry in two-dimensional quantum field theory'',
Nucl . Phys. {\bf B241} 333 (1984).
\bibitem{AdS/CFT} J. M. Maldacena, ``The large N limit of superconformal
field theories and supergravity'', Adv. Theor. Math. Phys. {\bf 2}
231 (1998) [{arXiv:hep-th/9711200}];\\S. S. Gubser, I. R. Klebanov
and A. M. Polyakov, ``Gauge theory correlators from non-critical
string theory'',
Phys. Lett. {\bf B428} 105 (1998) [{arXiv:hep-th/9802109}];\\
E. Witten, ``Anti-de Sitter space and holography'', Adv. Theor.
Math. Phys. {\bf 2} 253 (1998) [{arXiv:hep-th/9802150}].

\bibitem{review} N. Beisert et.al., ``Review of AdS/CFT Integrability: An Overview'',
To appear in Lett. Math. Phys. [arXiv:hep-th/1012.3982].

\bibitem{SpecCurv} A. Kazakov, A. Marshakov, J. A. Minahan and K. Zarembo,
``Classical/quantum integrability in AdS/CFT'', JHEP {\bf 0405} 024 (2004)
[arXiv:hep-th/0402207].

\bibitem{KRT} M. Kruczenski, J. Russo and A. A. Tseytlin, ``Spiky strings and giant
magnons on S5'', JHEP {\bf 0610} 002 (2006) [{arXiv:hep-th/0607044}].

\bibitem{sugra}
D. Z. Freedman, S. D. Mathur, A. Matusis and L. Rastelli, ``Correlation
functions in the CFTd/AdSd+1 correspondence'', Nucl. Phys. {\bf B546} 96 (1999) 96
[arXiv:hep-th/9804058];\\
G. Chalmers, H. Nastase, K. Schalm and R. Siebelink, ``R-current correlators in N =
4 super Yang-Mills theory from anti-de Sitter supergravity'', Nucl. Phys. {\bf B540} 247 (1999) [arXiv:hep-th/9805105];\\
S. Lee, S. Minwalla, M. Rangamani and N. Seiberg, ``Three-point functions of chiral
operators in D = 4, N = 4 SYM at large N'', Adv. Theor. Math. Phys. {\bf 2}
697 (1998) [arXiv:hep-th/9806074];\\
G. Arutyunov and S. Frolov, ``Some cubic couplings in type IIB supergravity on AdS5× S5 and three-point functions in SYM(4) at large N'', Phys. Rev. {\bf D61} 064009
(2000) [arXiv:hep-th/9907085].
\bibitem{Buchbinder}
E. I. Buchbinder, ``Energy-Spin Trajectories in AdS5 × S5 from Semiclassical Vertex
Operators'', JHEP {\bf 1004} 107 (2010) [arXiv:hep-th/1002.1716].
\bibitem{Janik} R. A. Janik, P. Surowka and A. Wereszczynski, ``On correlation functions of operators dual to classical spinning string states'', JHEP
{\bf 1005} 030 (2010) [arXiv:hep-th/1002.4613].
\bibitem{BuchTsey} E. I. Buchbinder and A. A. Tseytlin, ``On semiclassical approximation for cor-relators of closed string vertex operators in AdS/CFT'',
JHEP {\bf 1008} 057 (2010) [arXiv:hep-th/1005.4516].
\bibitem{Zarembo} K. Zarembo, ``Holographic three-point functions of semiclassical states'', JHEP {\bf 1009} 030 (2010) [arXiv:hep-th/1008.1059].
\bibitem{Costa} M. S. Costa, R. Monteiro, J. E. Santos and D. Zoakos,
``On three-point correlation functions in the gauge/gravity duality'',
JHEP {\bf 1011} 141 (2010) [arXiv:hep-th/1008.1070].
\bibitem{Roiban} R. Roiban and A. A. Tseytlin, ``On semiclassical computation of 3-point functions of closed string vertex operators in $AdS_5\times S^5$'',
Phys. Rev. {\bf D82} 106011 (2010) [arXiv:hep-th/1008.4921].
\bibitem{Hernandez} R. Hern\'andez, ``Three-point correlation functions from semiclassical circular strings'', J. Phys. {\bf A44} 085403 (2011)
[arXiv:hep-th/1011.0408].
\bibitem{Ryang} S. Ryang, ``Correlators of Vertex Operators for Circular Strings with Winding Numbers in $AdS_5\times S^5$'', JHEP {\bf 1101} 092 (2011)  [arXiv:hep-th/1011.3573].
\bibitem{Rashkov} D. Arnaudov and R. C. Rashkov, ``On semiclassical calculation of three-point functions
in $AdS_4\times CP^3$'', Phys. Rev. {\bf D83} 066011 (2011) [arXiv:hep-th/1011.4669].
\bibitem{Georgiou} G. Georgiou, ``Two and three-point correlators of operators dual to folded string solutions at strong coupling'', 
JHEP {\bf 1102} 046 (2011)  [arXiv:hep-th/1011.5181].
\bibitem{Russo} J. G. Russo and A. A. Tseytlin, ``Large spin expansion of semiclassical 3-point correlators in $AdS_5\times S^5$'', JHEP {\bf 1102} 029 (2011)  [arXiv:hep-th/1012.2760].
\bibitem{Park} C. Park and B. H. Lee, ``Correlation functions of magnon and spike'' [arXiv:hep-th/1012.3293];\\
 X. Bai, B. H. Lee and C. Park, ``Correlation function of dyonic strings''
[arXiv:hep-th/1104.1986].
\bibitem{Bak} D. Bak, B. Chen and J. B. Wu, ``Holographic Correlation Functions for Open Strings and Branes'',
JHEP {\bf 1106} 014 (2011) [arXiv:hep-th/1103.2024].
\bibitem{Bissi} A. Bissi, C. Kristjansen, D. Young and K. Zoubos, ``Holographic three-point functions of giant gravitons'',
JHEP {\bf 1106} 085 (2011) [arXiv:hep-th/1103.4079].
\bibitem{Arnaudov} D. Arnaudov, R. C. Rashkov and T. Vetsov, Three- and four-point correlators of operators dual to folded string solutions in $AdS_5\times S^5$'' [arXiv:hep-th/1103.6145].
\bibitem{BuchbTseyt} E. I. Buchbinder and A. A. Tseytlin, ``Semiclassical four-point functions in $AdS_5\times S^5$'', JHEP {\bf 1102} 072 (2011) [arXiv:hep-th/1012.3740].

\bibitem{Hernandez2} R. Hern\'andez, ``Three-point correlators for giant magnons'',
JHEP {\bf 1105} 123 (2011) [arXiv:hep-th/1104.1160].

\bibitem{Gromov}  J. Escobedo, N. Gromov, A. Sever and P. Vieira, ``Tailoring
Three-Point Functions and Integrability'' [arXiv:hep-th/1012.2475].

\bibitem{Vieira}  J. Escobedo, N. Gromov, A. Sever and P. Vieira, ``Tailoring
Three-Point Functions and Integrability II. Weak/strong coupling match''
[arXiv:hep-th/1104.5501].


\bibitem{AFZ06} G. Arutyunov, S. Frolov, M. Zamaklar, ``Finite-size Effects from Giant
Magnons'',  Nucl. Phys. \textbf{B778} 1 (2007) [arXiv:hep-th/0606126].

\bibitem{AFGS} D. Astolfi, V. Forini, G. Grignani, and G.W. Semenoff,
`` Gauge invariant finite size spectrum of the giant magnon '',
Phys. Lett. {\bf B651} 329 (2007) [arXiv:hep-th/0702043]

\bibitem{KM0803} T. Klose and T. McLoughlin, ``Interacting finite-size magnons'',
J. Phys. {\bf A41} 285401 (2008) [arXiv:hep-th/0803.2324v2]

\bibitem{AB1} C. Ahn and P. Bozhilov, ``Finite - size Effects for Single
Spike'', JHEP {\bf 0807} 105 (2008) [arXiv:hep-th/0806.1085v3].

\bibitem{HM06} D.M. Hofman and J. Maldacena,
``Giant magnons'', J. Phys. {\bf A39} 13095-13118 (2006)
[arXiv:hep-th/0604135v2]


\bibitem{TBA} D. Bombardelli, D. Fioravanti and R. Tateo, ``Thermodynamic
Bethe Ansatz for planar AdS/CFT: a proposal'',
J. Phys. {\bf A42} 375401 (2009) [arxiv:hep-th/0902.3930];\\
N. Gromov, V. Kazakov, A. Kozak and P. Vieira, ``Exact Spectrum of Anomalous Dimensions of Planar N = 4 Supersymmetric Yang-Mills Theory: TBA and excited states'',
Lett. Math. Phys. {\bf 91} 265 (2010) [arxiv:hep-th/0902.4458];\\
G. Arutyunov and S. Frolov, ``Thermodynamic Bethe Ansatz for the
$AdS_5\times S^5$ Mirror Model'', JHEP {\bf 0905} 068 (2009) [arxiv:hep-th/0903.0141].

\bibitem{JanLuk} R. A. Janik and T. Lukowski, ``Wrapping interactions at strong coupling - the giant magnon'', Phys. Rev. {\bf D76} 126008
(2007) [arXiv:hep-th/0708.2208].

\bibitem{HatSuz} Y. Hatsuda and R. Suzuki, ``Finite-size effects from giant magnons'', Nucl. Phys. {\bf B800} 349 (2008) [arXiv:hep-th/0801.0747].


\bibitem{Mpaper} C. Ahn and P. Bozhilov, ``Finite-size effects of membranes on 
$AdS_4 \times S^7$'', JHEP {\bf 0808} 054 (2008)  [arXiv:hep-th/0807.0566].

\bibitem{PB10} P. Bozhilov, ``Close to the Giant Magnons''
[arXiv:hep-th/1010.5465v1]



\end{thebibliography}
\end{document}